\documentstyle[preprint,aps,floats,epsfig]{revtex}

\begin{document}

\draft

\preprint{\rightline{ANL-HEP-PR-01-009}}

\title{Scaling behaviour at the $N_t=6$ chiral phase transition for
      2-flavour lattice QCD with massless staggered quarks, and an irrelevant
      4-fermion interaction.}

\author{J.~B.~Kogut}
\address{Dept. of Physics, University of Illinois, 1110 West Green Street,
Urbana, IL 61801-3080, USA}
\author{D.~K.~Sinclair}
\address{HEP Division, Argonne National Laboratory, 9700 South Cass Avenue,
Argonne, IL 60439, USA}

\maketitle
\begin{abstract}
We have simulated lattice QCD with 2 flavours of massless staggered quarks.
An irrelevant chiral 4-fermion interaction was added to the standard quark
action to allow us to simulate at zero quark mass. Thermodynamics was
studied on lattices of temporal extent 6. Clear evidence for a second order
chiral transition was observed and the critical exponents $\beta_{mag}$, 
$\delta$, $\nu$ and $\gamma_{mag}$ were measured. These exponents did not
agree with those expected by standard universality arguments. They were,
however, consistent with tricritical behaviour. The $\pi$ and $\sigma$
screening masses were measured and showed clear evidence for chiral symmetry
restoration at this transition.
\end{abstract}


\pagestyle{plain}
\parskip 5pt
\parindent 0.5in

\narrowtext

\section{Introduction}

In standard methods for simulating lattice QCD, \cite{hybrid}
the Dirac operator
(matrix) becomes singular in the limit of zero quark mass. Since these methods
require inversion of the Dirac operator they therefore fail for zero quark
mass. The iterative methods for performing these inversions become
progressively less efficient, requiring a number of iterations which diverges
as this zero mass limit is approached. This means that, even at the mass of
the physical $u$ and $d$ quarks, simulations on modest size lattices become
prohibitively expensive in computer time, and most simulations are performed
with unphysically large light-quark masses.

QCD with 2 light quark flavours is expected to undergo a second order
chiral transition at finite temperature, only in the zero mass limit. It would
thus be very helpful to be able to simulate in this limit in order to extract
detailed information about this transition such as critical exponents and the
equation of state in the neighbourhood of this transition. Here we note a
second difficulty with the standard action. In the zero mass limit on a finite
lattice, the chiral condensate which we need to use as an order parameter
vanishes on a configuration by configuration basis, which restricts how small
a mass we can use on a given lattice size. This contrasts with the behaviour of
most lattice spin models, where the magnetization (the order parameter for this
class of models) has a finite value close to the infinite volume limit (for
large enough lattices), even for zero magnetic field, for each configuration.
The way in which the magnetization averages to zero over the ensemble of
configurations, which it must for zero applied magnetic field, is that the
direction of magnetization changes from configuration to configuration to
enforce this. On a large enough lattice, a reasonable approximation to the
spontaneous magnetization is obtained by averaging the magnitude of the
magnetization (or its square) over the ensemble.

To enable simulations at zero quark mass, we have modified the standard
staggered quark action by the addition of an irrelevant 4-fermion term which
preserves the symmetries of the lattice action \cite{lks}.
Such a term will not affect
the continuum limit of the theory. It does, however, serve to render the
Dirac operator non-singular at zero quark mass. In the hadronic phase, where
the chiral symmetry is spontaneously broken, this is easy to understand. The
chiral condensate provides an effective mass to the quarks through the 4-fermion
coupling. Even where the 4-fermion coupling is small, as it must be close to
the continuum limit, this ``constituent'' quark mass is large enough to keep
the number of iterations required to invert the Dirac operator manageable. With
the standard staggered fermion action, the finite time extent of the lattice
keeps the Dirac operator from becoming singular in the plasma phase . It does, 
however, become 
nearly singular for topologically non-trivial gauge configurations due to
the lattice remnant of the Atiyah-Singer index theorem \cite{sv}. 
For our modified action,
the conservation of the axial quark-number current is explicitly broken by
the 4-fermion interaction, thus avoiding this consequence of the index theorem.
This keeps the number of iterations required to invert the Dirac operator
manageable in the plasma phase.

Although early attempts to extract critical exponents from staggered-quark
lattice QCD appeared promising \cite{karsch},
later attempts showed significant deviations
from the $O(4)$ or $O(2)$ exponents expected from universality arguments.
\cite{aoki,boyd}
Attempts to fit scaling with $\beta=6/g^2$ and quark mass $m$ to the scaling
functions of the $O(4)$ and $O(2)$ 3-dimensional spin models \cite{toussaint}
failed for lattices
of temporal extent $N_t=4$ \cite{milc,laermann}. 
While the situation appeared better for $N_t > 6$
($N_t = 6$ was somewhat ambiguous) insufficient ``data'' existed to make this
conclusive.

Our earlier simulations with our new action on $N_t=4$ lattices at $m=0$
indicated that the transition was probably first order \cite{lks}. 
Ongoing simulations
on an $N_t=4$ lattice with a smaller 4-fermion coupling indicate that the
chiral transition is either a first order or a very narrow second order
transition. The results of our simulations on $12^3 \times 6$ and 
$18^3 \times 6$ lattices summarized in an earlier letter \cite{ks} 
and presented in
detail in this paper, indicate that the $N_t=6$ transition is second order,
but with the critical exponents of the 3-dimensional tricritical point, rather
that those of the $O(4)$ or $O(2)$ spin models. We were able to arrive at this
conclusion because we were able to study the scaling of the chiral condensate
with $\beta$ at $m=0$ and thus extract the critical exponent $\beta_{mag}$.

Section 2 describes our modified action. In section 3 we discuss the critical
phenomena relevant to these simulations. We describe our simulations and
results in section 4. In section 5 we discuss our fits to critical scaling.
Section 6 gives our conclusions.

\section{Lattice QCD with chiral 4-fermion interactions.}

We have modified our staggered quark lattice QCD action by the addition of
an irrelevant chiral 4-fermion interaction. In the continuum the Euclidean
Lagrangian density of this theory is
\begin{equation}
{\cal L}=\frac{1}{4}F_{\mu\nu}F_{\mu\nu}
        +\bar{\psi}(D\!\!\!\!/+m)\psi
        -{\lambda^2 \over 6 N_f}[(\bar{\psi}\psi)^2
                          -(\bar{\psi}\gamma_5\tau_3\psi)^2].
\label{eqn:lagrangian}
\end{equation}
Lattice field theories incorporating fermions interacting both through gauge
fields and through quartic self-interactions have been studied before --- see
for example  \cite{kmy}, \cite{brower}, \cite{hands1,kim,hands2} ---  with
promising preliminary results.

Equation~\ref{eqn:lagrangian} can be rendered quadratic in the fermion fields
by the standard trick of introducing (non-dynamical) auxiliary fields $\sigma$ 
and $\pi$ in terms of which this Lagrangian density becomes
\begin{equation}
{\cal L}=\frac{1}{4}F_{\mu\nu}F_{\mu\nu}
        +\bar{\psi}(D\!\!\!\!/+\sigma+i\pi\gamma_5\tau_3+m)\psi
        +{3 N_f \over 2 \lambda^2}(\sigma^2+\pi^2)
\end{equation}
The molecular dynamics Lagrangian for a particular staggered fermion lattice
transcription of this theory in which $\tau_3$ is identified with $\xi_5$,
the flavour equivalent of $\gamma_5$, is
\begin{eqnarray}
L & = & -\beta\sum_{\Box}[1-\frac{1}{3}{\rm Re}({\rm Tr}_{\Box} UUUU)]
        +{N_f \over 8}\sum_s \dot{\psi}^{\dag} A^{\dag} A\dot{\psi}
        -\sum_{\tilde{s}}\frac{1}{8}N_f\gamma(\sigma^2+\pi^2)  \nonumber \\
  &   & +\frac{1}{2}\sum_l(\dot{\theta}_7^2+\dot{\theta}_8^2
        +\dot{\theta}_1^{\ast}\dot{\theta}_1
        +\dot{\theta}_2^{\ast}\dot{\theta}_2
        +\dot{\theta}_3^{\ast}\dot{\theta}_3)
        +\frac{1}{2}\sum_{\tilde{s}}(\dot{\sigma}^2+\dot{\pi}^2)
\label{eqn:hybrid}
\end{eqnarray}
where
\begin{equation}
A = \not\!\! D + m + \frac{1}{16} \sum_i (\sigma_i+i\epsilon\pi_i)
\end{equation}
with $i$ running over the 16 sites on the dual lattice neighbouring the site
on the normal lattice, $\epsilon=(-1)^{x+y+z+t}$ and $\not\!\! D$ the usual
gauge-covariant ``d-slash'' for the staggered quarks. The factor ${N_f \over 8}$
in front of the pseudo-fermion kinetic term is appropriate for the hybrid
molecular dynamics algorithm with ``noisy'' fermions, where $A\dot{\psi}$ are
chosen from a complex gaussian distribution with width 1. The ``dots''
represent derivatives with respect to molecular dynamics ``time'' as distinct
from normal time. We note that $\gamma=3/\lambda^2$. Although the determinant
of $A$ does not appear to be real, it becomes so in the continuum limit.
Without the gauge fields, this theory reverts to the one studied in
\cite{kim}, with $3 N_f$ flavours.

The advantage of this choice of the chiral 4-fermion interaction is that it 
preserves the axial $U(1)$ chiral symmetry of the normal staggered quark
lattice QCD action generated by $\gamma_5\xi_5$ at $m=0$. This means that,
when chiral symmetry is spontaneously broken, the pion associated with
$\xi_5\gamma_5$ will be a true Goldstone boson and will be massless at $m=0$,
even for finite lattice spacing. Under this exact chiral symmetry the fields
transform as
\begin{eqnarray}
\dot{\psi}(n) & \rightarrow & e^{-i\frac{1}{2}\phi\epsilon(n)}\dot{\psi}(n) \\
\sigma(n)+i\pi(n) & \rightarrow & e^{i\phi}[\sigma(n)+i\pi(n)]
\label{eqn:chiral}
\end{eqnarray}
from which we find that
\begin{eqnarray}
A\dot{\psi}(n) & \rightarrow & e^{i\frac{1}{2}\phi\epsilon(n)}A\dot{\psi}(n) \\
\sigma(n)+i\epsilon(n)\pi(n) & \rightarrow & e^{i\phi\epsilon(n)}
                                               [\sigma(n)+i\epsilon(n)\pi(n)],
\end{eqnarray}
when $m=0$. Hence, for massless quarks the above Lagrangian has an exact $U(1)$
axial flavour symmetry, the same remnant of continuum chiral flavour symmetry
as the standard staggered lattice Lagrangian.

\section{Critical behaviour of the chiral transition in lattice QCD}

Near a critical point, the order parameters of the theory in question scale
as powers of the parameters in the theory. Since chiral symmetry is restored
at the finite temperature transition from hadronic matter to a quark-gluon
plasma, the appropriate order parameter is the chiral condensate,
$\langle\bar{\psi}\psi\rangle$. The equation of state describing the scaling
of this order parameter close to the critical point is of the form \cite{amit}
\begin{equation}
\langle\bar{\psi}\psi\rangle = h^{1/\delta}f_{QCD}(th^{-1/\beta_{mag}\delta})
\label{eqn:eos}
\end{equation}
where $h = m/T$ and $t = \beta - \beta_c \propto (T-T_c)/T_c$. $f_{QCD}$ is
called the scaling function. This also serves to define 2 of the critical
exponents, $\beta_{mag}$ and $\delta$. Since $\langle\bar{\psi}\psi\rangle$ is
non-zero for $h=0$, $t < 0$ equation~\ref{eqn:eos} requires 
\begin{equation}
f_{QCD}(x) = C (-x)^\beta\,;  \;\;\;\;\;\; x \rightarrow -\infty,
\end{equation}
so that
\begin{equation}
\langle\bar{\psi}\psi\rangle = C (-t)^{\beta_{mag}}; \;\;\;\;\;\; t <0.
\end{equation}
At $t=0$,
\begin{equation}
\langle\bar{\psi}\psi\rangle = D h^{1/\delta}; \;\;\;\;\;\; D = f_{QCD}(0).
\end{equation}
Clearly, since $\langle\bar{\psi}\psi\rangle = \gamma \langle\sigma\rangle$,
similar scaling relations exist for $\langle\sigma\rangle$. In addition, at
$h=0$, the $\sigma$ screening mass vanishes as
\begin{equation}
m_\sigma = A |t|^\nu,
\end{equation}
since the correlation length $\xi$ of the system diverges as $\xi \propto
|t|^{-\nu}$. Finally we need to consider the scaling relations for the
susceptibilities which measure the fluctuations in the order parameter(s). For
reasons which will be explained later, we consider the scaling of the $\sigma$
susceptibility ($\chi_\sigma$), rather than that for
$\langle\bar{\psi}\psi\rangle$, with $t$. Near $t=0$, $\chi_\sigma$ diverges as
\begin{equation}
\chi_\sigma = V \left[\langle\langle\sigma\rangle^2\rangle-
\langle\langle\sigma\rangle\rangle^2\right] = c |t|^{-\gamma_{mag}}.
\end{equation}
Here the inner angle brackets represent averages over the lattice for a single 
configuration while the outer angle brackets represent averages over the
ensemble of configurations.

Let us briefly give the arguments which indicate the expected universality
class for this transition. First one notes that the Matsubara frequencies
for bosonic excitations are $2n\pi T$, while those for fermionic excitations
are $(2n+1)\pi T$ \cite{kapusta}. 
The expectation that only low lying excitations contribute
to critical phenomena leads one to the expectation that, at high T, only the
lowest bosonic excitation $n=0$ contributes and the fermions decouple. Since
this lowest bosonic mode is independent of time, this leads to an effective
3-dimensional theory (dimensional reduction) \cite{pw}. 
Since the masses of the 
$(\pi,\sigma)$ chiral multiplet vanish at the chiral transition, these
excitations alone should determine the critical scaling at this transition.
The effective field theories describing the interactions of the $(\pi,\sigma)$
multiplet are 3-dimensional $O(4)$ sigma models. Thus 2-flavour QCD should
lie in the universality class of the $O(4)$ sigma model in 3-dimensions 
\cite{pw}, with 
critical exponents $\beta_{mag} = 0.384(5)$, $\delta = 4.85(2)$, 
$\nu = 0.749(4)$ and $\gamma_{mag} = 1.471(6)$ \cite{spin}. 
Of course, lattice QCD with staggered quarks has only
one Goldstone pion, and this symmetry is reduced to $O(2)$. Thus if the
lattice is coarse enough that this symmetry breaking is relevant, 2-flavour
QCD should lie in the universality class of the $O(2)$ sigma model in
3-dimensions, with critical exponents $\beta_{mag} = 0.35(1)$, $\delta =
4.81(1)$, $\nu = 0.679(3)$ and $\gamma_{mag} = 1.328(6)$ \cite{spin}.

Of course, other possibilities exist. First the transition could be first
order, which appears to be the case at $N_t=4$ with our first choice of
4-fermi coupling \cite{lks}. 
Once the lattice spacing is large enough that lattice
artifacts can affect the transition, more symmetry breaking and symmetry
conserving operators come into play giving a more complex transition. The
simplest of these is the tricritical point. The sigma models which give the
universal critical points discussed in the previous paragraph lie in the
universality class of the polynomial sigma models which are quartic in the
fields. Tricritical behaviour occurs in sigma models which are sixth order in
the fields \cite{dl-ls}. 
This means that there is a second symmetry preserving parameter or
temperature and a second symmetry breaking parameter or magnetic field. Since
3 is the upper critical dimension for such sigma models, the critical exponents
are given by a mean field analysis. Thus $\beta_{mag} = \frac{1}{4},
\frac{1}{2}$, the $\delta = 5, 3$, $\nu = \frac{1}{2}$, $\gamma=1,
\frac{1}{2}$. The 2 values of $\beta_{mag}$, $\delta$ and $\gamma$ describe
the scalings when the 2 `magnetic fields' are varied independently. Even more
values are possible when we take into account the second `temperature'. A
tricritical point occurs where a line of first order phase transitions
continues as a line of second order phase transitions or in general where
there are 2 competing phase transitions.

\section{Finite temperature simulations and results for $N_t=6$.}

Our $N_t=4$ simulations at $\gamma=10$ which showed evidence for a first order
transition were reported in an earlier paper \cite{lks}. 
Here we present the details of
our simulations on $12^3 \times 6$ and $18^3 \times 6$ lattices. A brief 
summary of these results were given in a letter \cite{ks}. 
Our $12^3 \times 6$ simulations were
performed at $\gamma=10$ and $\gamma=20$ while our $18^3 \times 6$ simulations
were done at $\gamma=20$.

We shall first discuss our results for zero mass quarks. Here, without the
benefit of the symmetry breaking interaction (quark mass term), the direction
of symmetry breaking changes over the run so that $\langle\sigma\rangle$ and
$\langle\pi\rangle$ and the corresponding chiral condensates
$\langle\bar{\psi}\psi\rangle$ and $\langle\bar{\psi}\gamma_5\xi_5\psi\rangle$
(where the $\langle\rangle$ is averaging only over the lattice, not the
ensemble) average to zero over the ensemble. It is in this way that the
absence of spontaneous breaking of this continuous symmetry on a finite
lattice is enforced. Note that this contrasts with the standard lattice
formulation, where if we could simulate at zero quark mass, the chiral
condensates would vanish on a configuration by configuration basis,
and there would be no way to predict the infinite lattice results. In our case,
we can estimate the infinite lattice results by averaging the magnitudes
$\sqrt{\langle\sigma\rangle^2 + \langle\pi\rangle^2}$ and
$\sqrt{\langle\bar{\psi}\psi\rangle^2-
\langle\bar{\psi}\gamma_5\xi_5\psi\rangle^2}$, introducing an error which is
of order $1/V$ below the phase transition and $1/\sqrt{V}$ above it, on a 
lattice of 4-volume $V$.

\begin{figure}[htb]
\epsfxsize=6in
\centerline{\epsffile{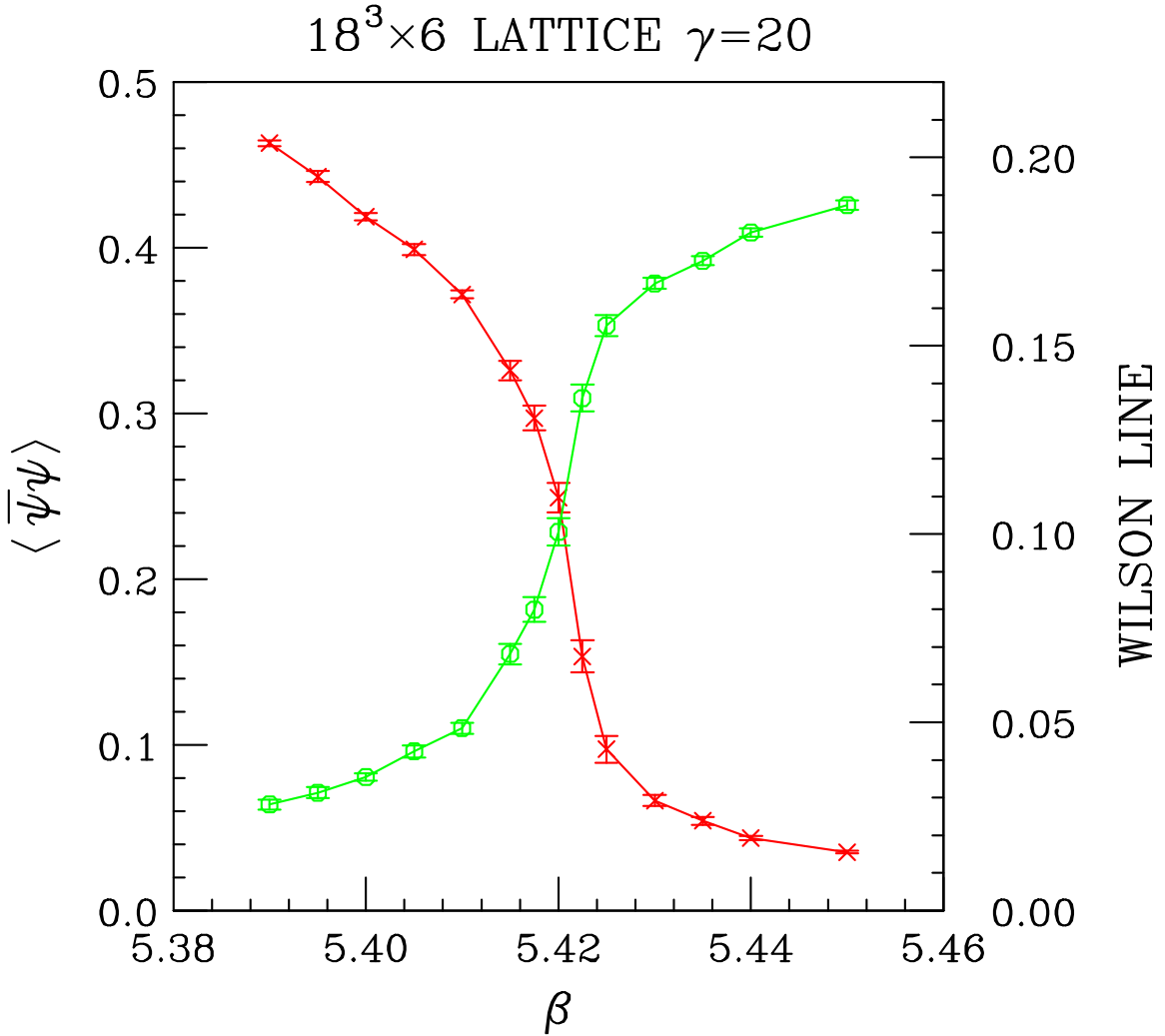}}
\caption{`$\langle\bar{\psi}\psi\rangle$' and Wilson line as functions of
$\beta$ on an $18^3 \times 6$ lattice at $\gamma=20$.\label{fig:wil-psi}}
\end{figure}

The simulations of equation~\ref{eqn:hybrid} were performed using the hybrid
molecular-dynamics algorithm with ``noisy'' fermions, allowing us to tune the
number of flavours to 2. The simulations on the $18^3 \times 6$ lattice were
performed at 14 $\beta$ values from 5.39 to 5.45. The behaviour of the chiral
condensate and Wilson line (Polyakov loop) are given in 
figure~\ref{fig:wil-psi}.
The values of the chiral condensate and $\langle\sigma\rangle$ (actually the
magnitudes defined above) as well as the length of each run in 
molecular-dynamics time units are given in table~\ref{tab:pbp-sigma18}. (Note
this ``time'' is twice that of equation~\ref{eqn:hybrid} for consistency with
that used by the HEMCGC and HTMCGC collaborations.)
\begin{table}[htb]
\begin{tabular}{lddc}
$\beta$ & $\langle\bar{\psi}\psi\rangle$  & $\langle\sigma\rangle$ & `Time' \\
\hline
5.3900  &  0.4630(18)  &  0.02543(10)  &  5000    \\
5.3950  &  0.4430(32)  &  0.02454(16)  &  5000    \\
5.4000  &  0.4186(23)  &  0.02337(11)  &  10000   \\
5.4050  &  0.3990(33)  &  0.02244(17)  &  10000   \\
5.4100  &  0.3719(23)  &  0.02109(12)  &  20000   \\
5.4150  &  0.3259(58)  &  0.01880(30)  &  40000   \\
5.4175  &  0.2973(74)  &  0.01734(40)  &  50000   \\
5.4200  &  0.2491(89)  &  0.01479(48)  &  50000   \\
5.4225  &  0.1535(96)  &  0.00957(53)  &  50000   \\
5.4250  &  0.0973(81)  &  0.00649(45)  &  20000   \\
5.4300  &  0.0664(33)  &  0.00479(18)  &  20000   \\ 
5.4350  &  0.0542(25)  &  0.00414(13)  &  10000   \\
5.4400  &  0.0439(12)  &  0.00364(7)   &  10000   \\
5.4500  &  0.0353(8)   &  0.00326(5)   &  5000    
\end{tabular}
\caption{$\langle\bar{\psi}\psi\rangle$, $\langle\sigma\rangle$ and run length
versus $\beta$ on an $18^3 \times 6$ lattice at $\gamma=20$.%
\label{tab:pbp-sigma18}}
\end{table}
Let us make several observations with regard to these measurements. The first is
that the phase transition occurs at $5.4225 \lesssim \beta_c \lesssim 5.4250$.
The second is that that due to fluctuations both
$\langle\bar{\psi}\psi\rangle$ and $\langle\sigma\rangle$ are appreciable in
the plasma phase. The third observation is that there are considerable
deviations from the expected relation
\begin{equation}
\langle\bar{\psi}\psi\rangle = \gamma\langle\sigma\rangle
\end{equation}
Part of this is attributable to these quantities having different finite volume
errors, which is why the deviations become worse as $\beta$ increases.
The second is ${\cal O}(dt^2)$ (and higher) errors which are quite large at
this $\gamma$ for the $dt=0.05$ we use. These errors affect
$\langle\sigma\rangle$ much more than $\langle\bar{\psi}\psi\rangle$ and could
presumably have been reduced by scaling the `kinetic' term in
equation~\ref{eqn:hybrid} by a scale factor greater than 1, thus slowing the
time evolution of $\sigma$ and $\pi$. For this reason we perform our critical
scaling analyses on $\langle\bar{\psi}\psi\rangle$ rather than
$\langle\sigma\rangle$. These assertions have been quantified for our $N_t=4$,
$\gamma=20$ simulations and are presented in table~\ref{tab:dt2}, which
suggests that both effects are coming into play. In fact, assuming a
$1/V$ dependence for $\langle\bar{\psi}\psi\rangle$ and
$\langle\sigma\rangle$ would give $\approx 0.7643$ and $\approx 0.03992$ for
their infinite volume limits on an $N_t=4$ lattice at $\beta=5.27$ and
$dt=0.05$, giving a ratio of $\approx 19.15$. If, in addition, we assumed the
coefficient of $1/V$ to be the same at $dt=0.0125$ we would find
$\langle\bar{\psi}\psi\rangle \approx 0.7567$ and $\langle\sigma\rangle
\approx 0.03797$ and the ratio $\approx 19.93$ at this $dt$. Assuming the
finite $dt$ error is ${\cal O}(dt^2)$ leads to an estimate of $\approx 19.98$
for the infinite volume zero $dt$ limit which is consistent with the correct
value of $20$.
\begin{table}[htb]
\begin{tabular}{lddd}
\multicolumn{4}{c}{$8^3 \times 4$ lattice} \\
$dt$  &  $\langle\bar{\psi}\psi\rangle$  &  $\langle\sigma\rangle$ &
$\langle\bar{\psi}\psi\rangle/\langle\sigma\rangle$ \\
\hline 
0.0125  &  0.7853(42)  &  0.04051(22)  &  19.38(4)  \\
0.0250  &  0.7898(42)  &  0.04114(21)  &  19.19(4)  \\
0.0500  &  0.7929(38)  &  0.04246(19)  &  18.68(4)  \\
0.1000  &  0.8175(30)  &  0.04715(16)  &  17.34(3)  \\
\hline
\multicolumn{4}{c}{$12^2 \times 24 \times 4$ lattice} \\
$dt$  &  $\langle\bar{\psi}\psi\rangle$  &  $\langle\sigma\rangle$ &        
$\langle\bar{\psi}\psi\rangle/\langle\sigma\rangle$ \\                      
\hline                                                                      
0.0500  &  0.7728(27)  &  0.04067(14)  &  19.00(3) \\ 
\end{tabular}
\caption{$dt$ and volume dependence of $\langle\bar{\psi}\psi\rangle$,
$\langle\sigma\rangle$ and their ratios at $\beta=5.27$.\label{tab:dt2}}
\end{table}
Our results for the $12^3 \times 6$ runs at $\gamma=20$ are presented in
table~\ref{tab:pbp-sigma12/20}, while those for our $12^3 \times 6$ runs
at $\gamma=10$ are given in table~\ref{tab:pbp-sigma12/10}.
\begin{table}[htb]
\begin{tabular}{lddc}
$\beta$ & $\langle\bar{\psi}\psi\rangle$  && `Time' \\
\hline
5.39000   &  0.478(3)  &&  3500   \\
5.39500   &  0.444(7)  &&  2000   \\
5.40000   &  0.427(3)  &&  3500   \\
5.40500   &  0.420(3)  &&  5000   \\
5.41000   &  0.376(4)  &&  6500   \\
5.41500   &  0.339(5)  &&  7500   \\
5.41750   &  0.299(6)  &&  40000  \\
5.42000   &  0.261(4)  &&  50000  \\
5.42125   &  0.245(7)  &&  30000  \\
5.42250   &  0.210(5)  &&  40000  \\
5.42500   &  0.200(5)  &&  45000  \\
5.43000   &  0.137(6)  &&  75000
\end{tabular}
\caption{$\langle\bar{\psi}\psi\rangle$ and run length
versus $\beta$ on an $12^3 \times 6$ lattice at $\gamma=20$.%
\label{tab:pbp-sigma12/20}}
\end{table}
\begin{table}[htb]
\begin{tabular}{lddc}
$\beta$ & $\langle\bar{\psi}\psi\rangle$  && `Time' \\
\hline
  5.4     &   0.648(5)  &&  1500   \\
  5.41    &   0.616(4)  &&  1500   \\
  5.42    &   0.585(5)  &&  1500   \\
  5.43    &   0.539(2)  &&  10000  \\
  5.44    &   0.496(2)  &&  10000  \\
  5.45    &   0.436(2)  &&  35000  \\
  5.455   &   0.389(2)  &&  70000  \\
  5.46    &   0.330(3)  &&  45000  \\
  5.46125 &   0.302(3)  &&  60000  \\
  5.4625  &   0.267(3)  &&  60000  \\
  5.46375 &   0.234(4)  &&  35000  \\
\end{tabular}
\caption{$\langle\bar{\psi}\psi\rangle$ and run length
versus $\beta$ on an $12^3 \times 6$ lattice at $\gamma=10$.%
\label{tab:pbp-sigma12/10}}
\end{table}

\begin{figure}[htb]
\epsfxsize=6in
\centerline{\epsffile{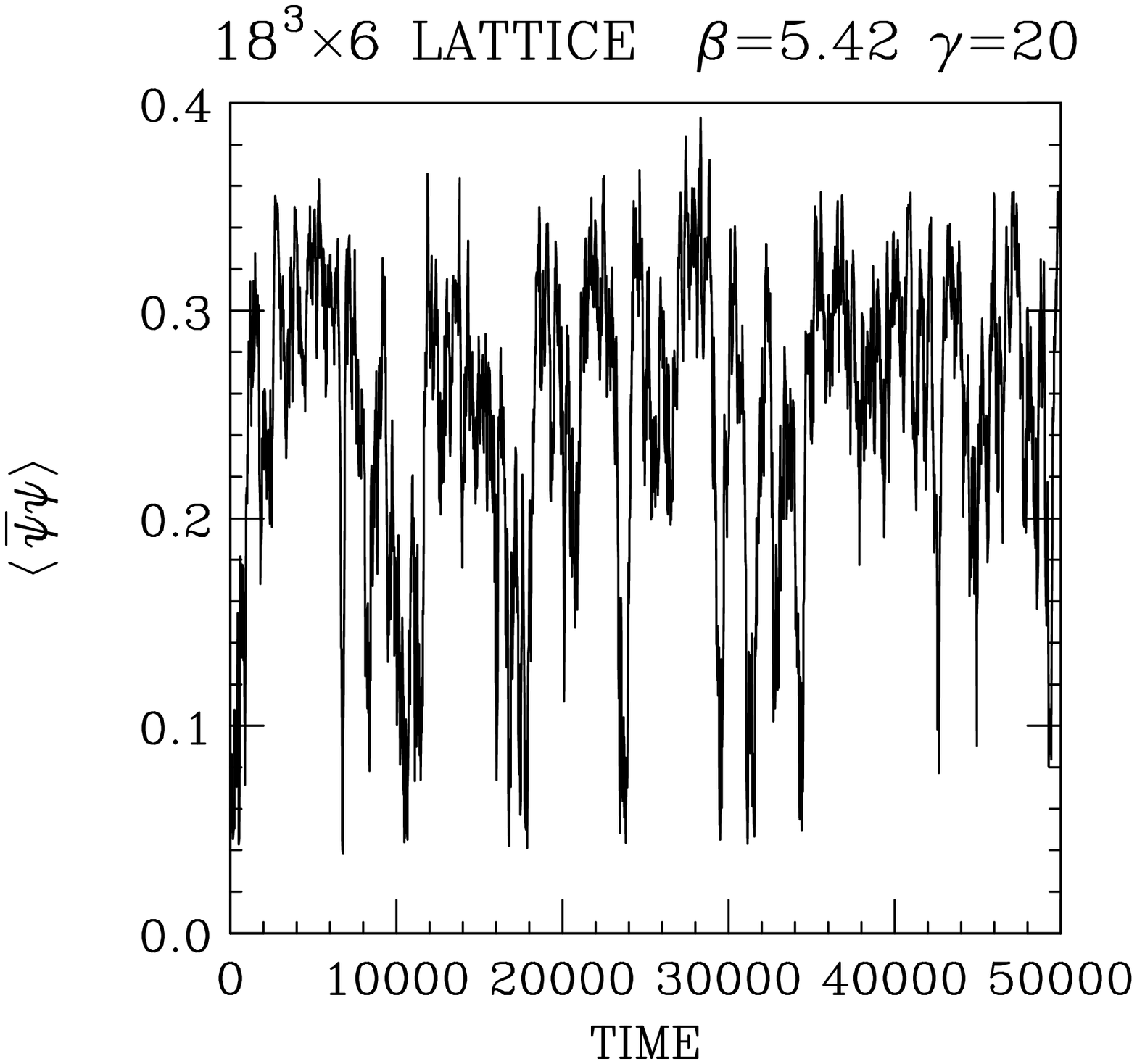}}
\caption{Time history of $\langle\bar{\psi}\psi\rangle$ at $\beta=5.42$.%
\label{fig:b542}}
\end{figure}

\begin{figure}[htb]
\epsfxsize=6in
\centerline{\epsffile{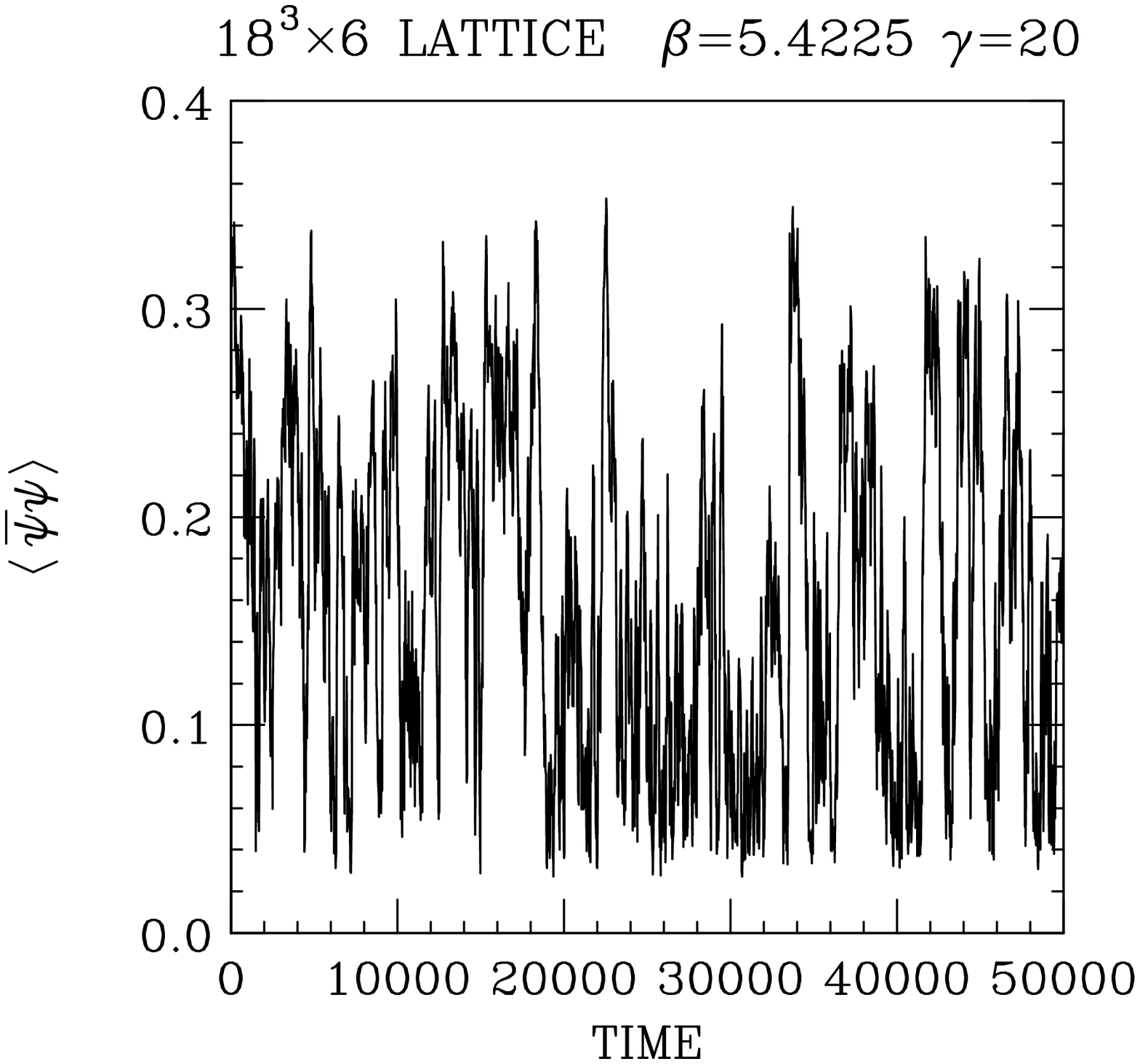}}
\caption{Time history of $\langle\bar{\psi}\psi\rangle$ at $\beta=5.4225$.%
\label{fig:b54225}}
\end{figure}

The smoothness of the transition with no hints of metastability suggests
a second order transition. Further evidence for this is the fluctuations
with long time constants which are obvious from time histories of the
chiral condensate close to the transition for the $18^3 \times 6$ lattice at
$\gamma=20$, shown in figures~\ref{fig:b542},~\ref{fig:b54225}.

\begin{figure}[htb]
\epsfxsize=6in
\centerline{\epsffile{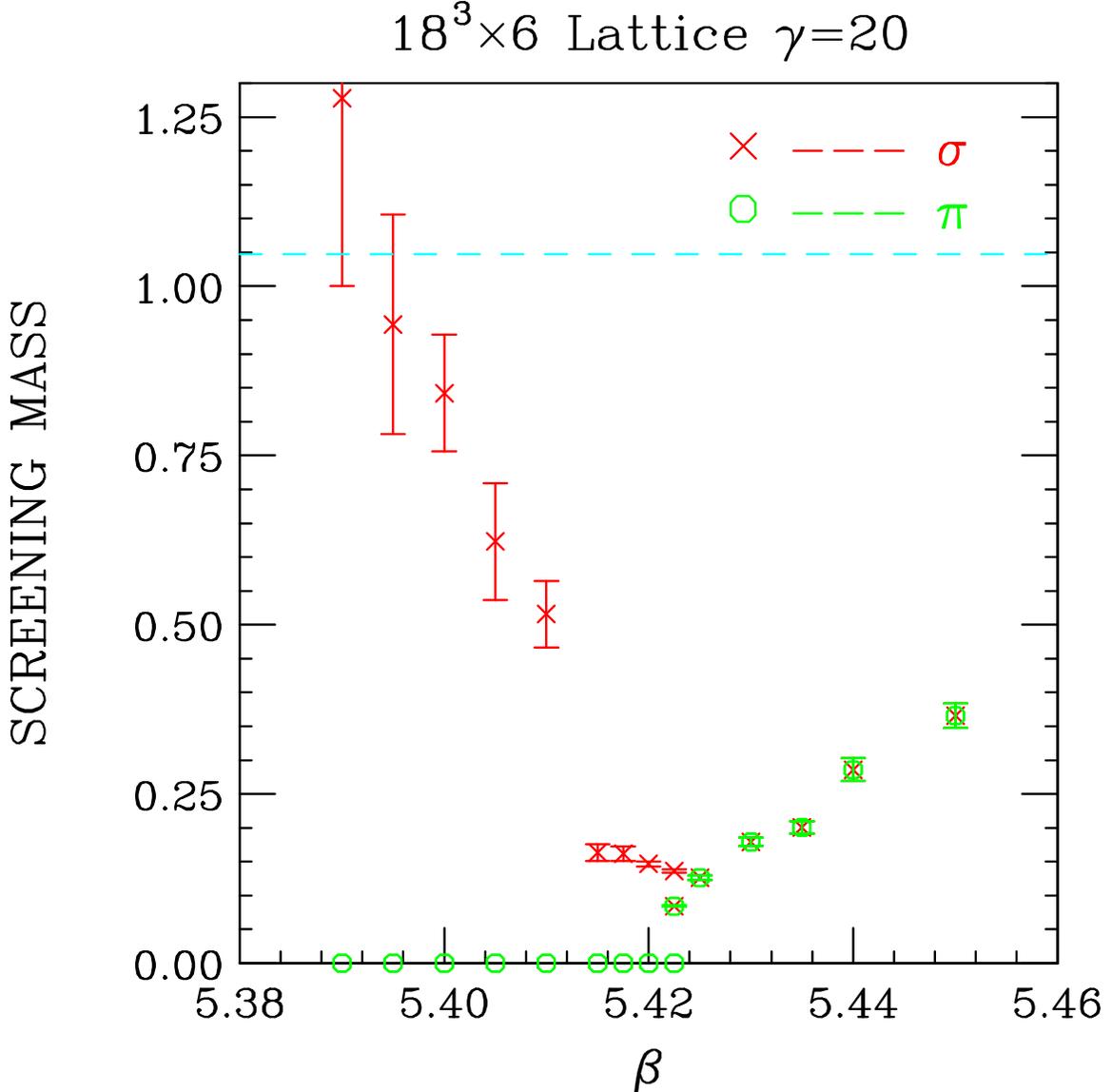}}
\caption{$\sigma$ and $\pi$ screening masses as functions of $\beta$. Note
that we have forced the the pion mass to be zero below the transition and
the $\sigma$ and $\pi$ masses to be equal above the transition. The dashed
line is at $\pi/3$.%
\label{fig:spmass}}
\end{figure}

On the largest ($18^3 \times 6$) lattice, we have measured the screening
propagators for the $\pi$ and $\sigma$ fields, i.e. the $\pi$ and $\sigma$
propagators for spatial separations. For this we use the $\pi$ and $\sigma$
auxiliary fields. This has the advantage over using fermion bilinears with
the same quantum numbers in that it automatically includes the disconnected
contributions to both these fields. Below the transition, we define our
$\sigma$ field by globally rotating our fields so that $\langle\pi\rangle=0$
on each configuration. This is the best we can do since the direction of
symmetry breaking rotates from configuration to configuration. Our zero momentum
$\sigma$ and $\pi$ screening propagators are then defined by
\begin{eqnarray}
S(Z) &=& \frac{1}{V}\sum_z\left[\sum_{xyt} \sigma(x,y,z,t)\sum_{x'y't'}
\sigma(x',y',z+Z,t')\right] - N_x N_y N_t\langle\sigma\rangle^2             \\
P(Z) &=& \frac{1}{V}\sum_z\left[\sum_{xyt} \pi(x,y,z,t)\sum_{x'y't'}
\pi(x',y',z+Z,t')\right]
\end{eqnarray}
We note that $P(Z)$, so defined, obeys
\begin{equation}
\sum_Z P(Z) = 0
\end{equation}
Hence we should fit P(Z) to the form
\begin{equation}
P(Z) \rightarrow A [exp(-Z m_\pi) + exp(-(N_z-Z) m_\pi)] + constant
\label{eqn:pifit}
\end{equation}
Such fits in this low temperature domain yield values of $m_\pi$ in the range
$0.1 < m_\pi < 0.3$. However, good fits are also obtained by fitting to a zero
mass pion propagator. The reason for this is clear. Over the relatively
small range we can fit these propagators on an $N_z=18$ lattice, 
equation~\ref{eqn:pifit} is well approximated by the parabola
\begin{equation}
P(Z) \rightarrow A m_\pi^2 (Z-N_z/2)^2 + [2 A exp(-m_\pi N_z/2) + constant]
\end{equation}
which is the form of the zero mass propagator. Hence when we graph the $\pi$
mass, we will use our knowledge that it must vanish when the chiral symmetry
is broken to set it to zero below the transition. We fit $S(Z)$ to the form
\begin{equation}                                                                
S(Z) \rightarrow B [exp(-Z m_\sigma) + exp(-(N_z-Z) m_\sigma)]
               + C [exp(-Z m_{\pi_2}) + exp(-(N_z-Z) m_{\pi_2})]             
\end{equation} 
and a similar fit with $C=0$, and choose the fit with the best signal in this
low-temperature phase. In the high temperature phase the restoration of chiral
symmetry means that we can no longer distinguish $\sigma$ and $\pi$. The best
propagator to use in this plasma phase is $PS(Z)$ defined as
\begin{equation}
PS(Z) = P(Z) + S(Z)
\end{equation}
Note here we do not have to rotate our fields. We then fit to the form
\begin{equation}
PS(Z) \rightarrow A [exp(-Z m_\pi) + exp(-(N_z-Z) m_\pi)].
\label{eqn:ps.mass}
\end{equation}
The screening masses obtained in this way are plotted in 
figure~\ref{fig:spmass}.
We note the $\sigma$ mass approaches zero as we approach the transition from
the low temperature side. The combined $\sigma$/$\pi$ mass rises from zero as
we enter the plasma phase. However, even at the highest $\beta$ and hence $T$
we consider, it is still far below the expected infinite temperature limit of
$2\pi T = \pi/3$.

\begin{figure}[htb]
\epsfxsize=6in
\centerline{\epsffile{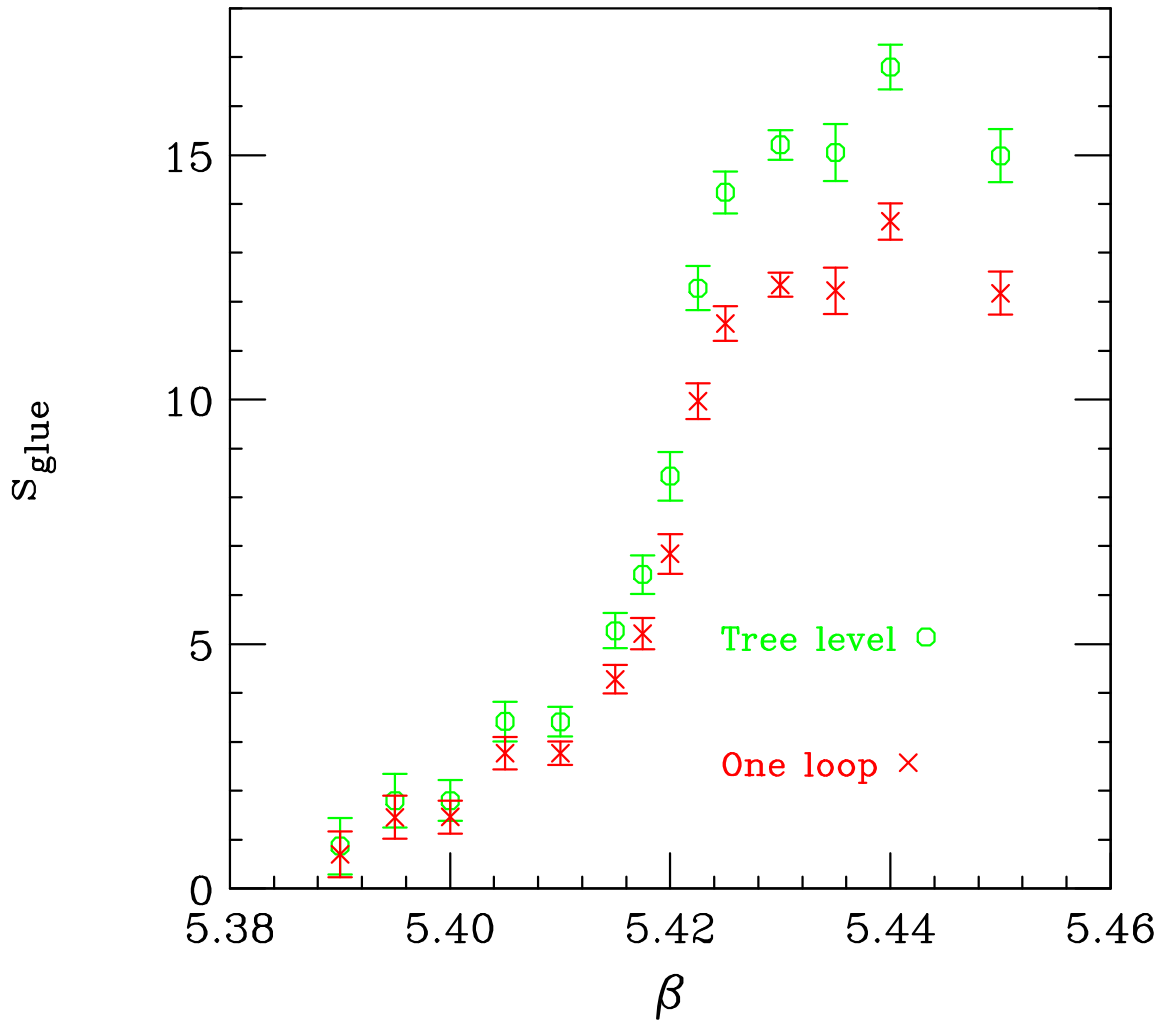}}
\caption{Gluon partial energy density as a function of $\beta$%
\label{fig:s_g}}
\end{figure}

\begin{figure}[htb]
\epsfxsize=6in
\centerline{\epsffile{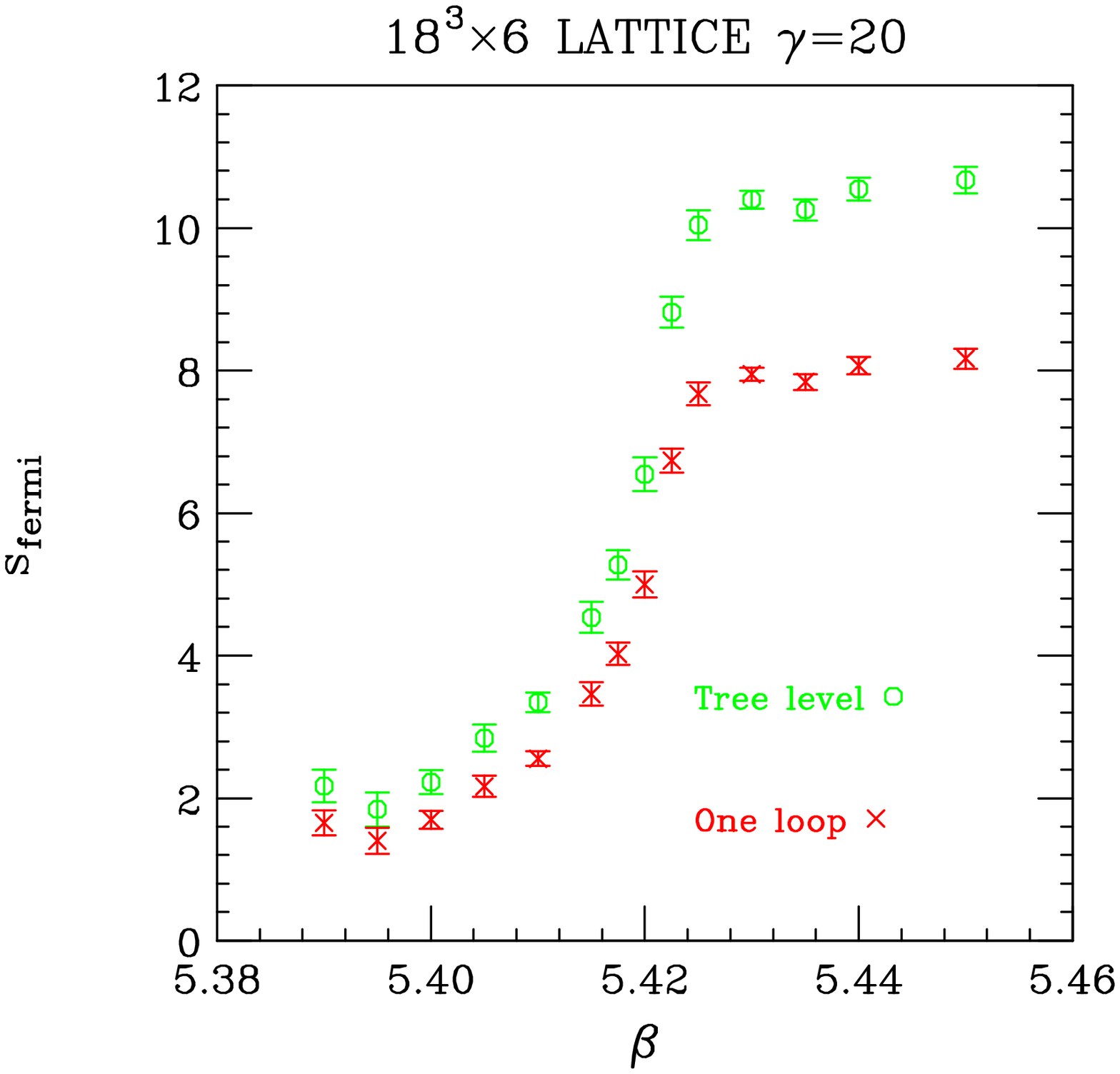}}       
\caption{Quark partial energy density as a function of $\beta$%
\label{fig:s_f}}   
\end{figure}

In addition, we measured contributions to the energy density, pressure and
entropy density. Extraction of energy density or pressure would require
measurements at the same $\beta$ values at zero temperature. Hence we 
concentrate on the entropy density which can be determined from our 
measurements. To 1-loop order the partial energy densities for the gluons and
the quarks are given by \cite{chavel}
\begin{equation}
{s_g \over T^3} = \frac{4}{3} N_t^4\frac{6}{g^2}\left[1+g^2\frac{1}{2}
                  \left(\frac{\partial C_\tau}{\partial\xi}-
                  \frac{\partial C_\sigma}{\partial\xi}\right)\right]
                  \left[\langle P_{ss}\rangle-\langle P_{st}\rangle\right]
\end{equation}
and
\begin{eqnarray}
{s_f \over T^3} = \frac{4}{3} N_t^4 && \left\{ \left(1+g^2\frac{\partial C_F}
{\partial\xi}\right)\frac{N_f}{4} \left[\bar{\psi}D\!\!\!\!/_0\psi 
-\frac{3}{4}\right]\right.      \nonumber                       \\
+ && \left.\frac{1}{2}\left(1+\frac{g^2}{2\pi^2}+4 g^2\frac{\partial C_m}
{\partial\xi}\right)\left[\gamma\frac{N_f}{8}\langle\sigma^2+\pi^2\rangle -1
\right]\right\}
\end{eqnarray}
where $P_{ss}$ and $P_{st}$ are the space-space and space-time plaquettes
respectively. The coefficient functions are given in reference~\cite{chavel}. 
Since such
expansions in terms of lattice couplings are notoriously poor, and an improved
staggered fermion lattice perturbation expansion \cite{golterman}
for these quantities has yet
to be performed, we present graphs of the tree level as well as the 1-loop
results for these quantities in figures~\ref{fig:s_g},~\ref{fig:s_f}. The
Stephan-Boltzmann value for $s_{g(lue)}/T^3$ is $32\pi^2/45=7.0184$. The finite
lattice size/spacing effects on an $18^3 \times 6$ lattice increase this to
$8.0951$. The Stephan-Boltzmann value for $s_{f(ermi)}/T^3$ is 
$7\pi^2 N_f/15=9.2116$ which becomes $16.9823$ on an $18^3 \times 6$ lattice.
We note that both partial entropies increase rapidly from rather small values
near the critical point and flatten out as the system passes into the plasma
phase. The gauge field partial entropy lies above the Stephan-Boltzmann value
in the plasma region, while the quark field partial entropy is about half
its Stephan-Boltzmann limit. However, it should be noted that, since the
$\pi/\sigma$ screening masses indicate that the quarks in these mesonic
excitations are still deeply bound at these $\beta$ values, the high temperature
limit would not be expected until a considerably higher $\beta$.
 
In addition to these zero mass simulations, we have performed simulations at
finite mass on a $12^3 \times 6$ lattice. For these simulations we fixed
$\beta=\beta_c$. Here, the presence of even a relatively small symmetry
breaking mass term was sufficient to align the condensate in the direction of
$\langle\bar{\psi}\psi\rangle$ (or $\langle\sigma\rangle$) so that we were able
to measure the chiral condensate directly. The chiral condensate at $\gamma=10$
as a function of quark mass is given in table~\ref{tab:pbp-m/10}, while that for
$\gamma=20$ is given in table~\ref{tab:pbp-m/20}.
\begin{table}[htb]
\begin{tabular}{ldc}
\multicolumn{3}{c}{$\gamma=10$,$\;\;\beta=5.465$}                     \\
$\beta$               &    $\langle\bar{\psi}\psi\rangle$  &  `Time'  \\
\hline
0.004                 &           0.397(3)         &   20000          \\
0.005                 &           0.427(3)         &   25000          \\
0.006                 &           0.446(2)         &   25000          \\
0.007                 &           0.461(2)         &   25000          \\
0.010                 &           0.512(2)         &   5000           \\
0.015                 &           0.561(1)         &   20000          \\
0.020                 &           0.606(2)         &   5000           \\
0.030                 &           0.674(2)         &   5000
\end{tabular}
\caption{Mass dependence of $\langle\bar{\psi}\psi\rangle$ at $\beta_c$ for
         $\gamma=10$.\label{tab:pbp-m/10}}
\end{table}
\begin{table}[htb]                                                              
\begin{tabular}{ldc}
\multicolumn{3}{c}{$\gamma=20$,$\;\;\beta=5.23$}                      \\
$\beta$              &    $\langle\bar{\psi}\psi\rangle$  &  `Time'   \\
\hline
0.0030               &              0.256(6)              &  10000    \\
0.0040               &              0.329(4)              &  10000    \\
0.0050               &              0.331(4)              &  10000    \\
0.0075               &              0.387(2)              &  10000    \\
0.0100               &              0.421(2)              &  15000    \\
0.0150               &              0.478(1)              &  15000    \\     
0.0200               &              0.534(3)              &  3000     \\
0.0300               &              0.594(2)              &  3000
\end{tabular}
\caption{Mass dependence of $\langle\bar{\psi}\psi\rangle$ at $\beta_c$ for
         $\gamma=20$.\label{tab:pbp-m/20}}
\end{table}

\section{Critical Scaling}

Much of what is presented here was presented in our letter \cite{ks}. 
We repeat these
results here for completeness, along with results which have not been presented
before.

Fitting the measured values of $\langle\bar{\psi}\psi\rangle$ on a $\gamma=20$, 
$18^3 \times 6$ lattice given in table~\ref{tab:pbp-sigma18} to the scaling
form $\langle\bar{\psi}\psi\rangle = C (\beta_c - \beta)^{\beta_{mag}}$ for the
interval $5.41 \leq \beta \leq 5.4225$ yields $\beta_c=5.4230(2)$, $C=1.19(12)$
and $\beta_{mag}=0.27(2)$ with a $94\%$ confidence level. On a $12^3 \times 6$
lattice, also at $\gamma=20$, fits to the values in 
table~\ref{tab:pbp-sigma12/20} gave $\beta_c=5.4249(7)$, $C=1.46(22)$ and
$\beta_{mag}=0.32(4)$ at a $27\%$ confidence interval. Our $12^3 \times 6$
measurements at $\gamma=10$ (table~\ref{tab:pbp-sigma12/10}) yielded 
$\beta_c=5.4650(1)$ and $\beta_{mag}=0.27(3)$.

\begin{figure}[htb]
\epsfxsize=6in
\centerline{\epsffile{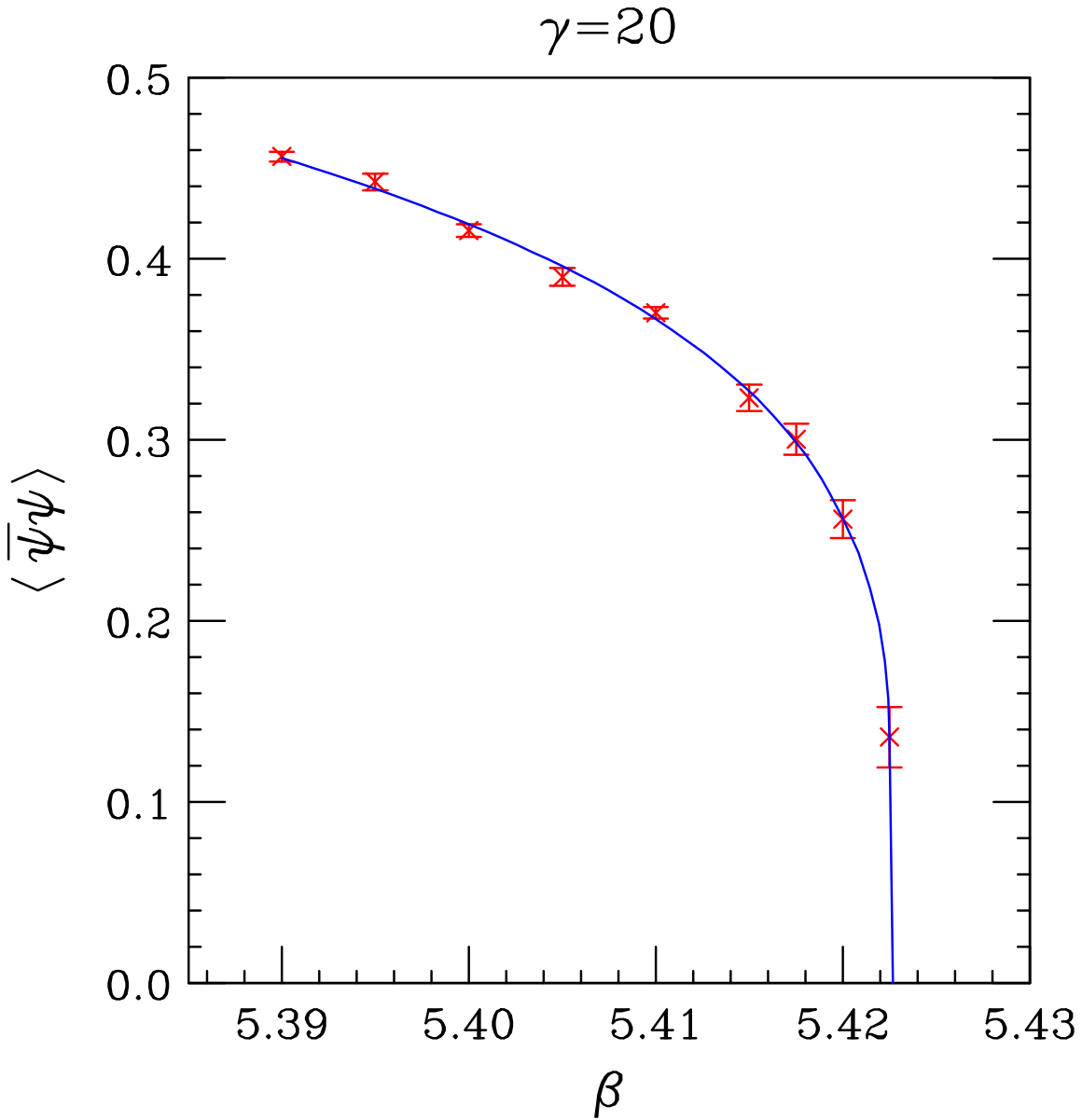}}
\caption{$\langle\bar{\psi}\psi\rangle$ versus $\beta$ with critical
scaling curve superimposed.\label{fig:pbp}}
\end{figure}

Clearly the sizable finite volume effects in these measurements are a major
source of systematic errors. Having two different size lattices at $\gamma=0$
enables us to remove the leading order finite volume effects. Here we note that
\begin{equation}
\langle\langle\bar{\psi}\psi\rangle^2-
\langle\bar{\psi}\gamma_5\xi_5\psi\rangle^2\rangle = 
\langle\langle\bar{\psi}\psi\rangle\rangle_\infty^2 + constant/V
\end{equation}
where here the inner $\langle\rangle$ stands for averaging over the lattice
and the outer $\langle\rangle$ stands for the ensemble average. 
Table~\ref{tab:pbp2} gives this quantity for the $12^3 \times 6$ lattice,
the $18^3 \times 6$ lattice and the extrapolation to an $\infty^3 \times 6$
lattice where it gives us an estimate of 
$\langle\langle\bar{\psi}\psi\rangle\rangle_\infty^2$.
\begin{table}[htb]
\begin{tabular}{lddd}
\multicolumn{4}{c}{$\langle\langle\bar{\psi}\psi\rangle^2-
                    \langle\bar{\psi}\gamma_5\xi_5\psi\rangle^2\rangle$}     \\
$\beta$   &   $12^3 \times 6$   &   $18^3 \times 6$  &   $\infty^3 \times 6$ \\
\hline
5.39000  &  0.231(2)   &  0.2150(17)   &  0.2083(25)     \\
5.39500  &  0.200(2)   &  0.1970(29)   &  0.1957(41)     \\
5.40000  &  0.184(2)   &  0.1760(19)   &  0.1726(28)     \\
5.40500  &  0.179(2)   &  0.1601(27)   &  0.1521(39)     \\
5.41000  &  0.145(1)   &  0.1393(16)   &  0.1370(24)     \\
5.41500  &  0.120(2)   &  0.1091(32)   &  0.1045(47)     \\
5.41750  &  0.0961(4)  &  0.0920(36)   &  0.0902(51)     \\
5.42000  &  0.0731(4)  &  0.0678(38)   &  0.0656(54)     \\
5.42125  &  0.0796(2)  &    ---        &    ---          \\
5.42250  &  0.0610(1)  &  0.0311(32)   &  0.0184(45)     \\
5.42500  &  0.0531(5)  &  0.0133(22)   & -0.0035(32)     \\
5.43000  &  0.0243(2)  &  0.00608(62)  & -0.0016(9)      \\
5.43500  &    ---      &  0.00396(41)  &    ---          \\
5.44000  &    ---      &  0.00255(17)  &    ---          \\
5.45000  &    ---      &  0.00160(7)   &    ---         
\end{tabular}
\caption{Averages of squares of the chiral condensate on $12^3 \times 6$
         and $18^3 \times 6$ lattices and the extrapolation to infinite
         spatial volume.\label{tab:pbp2}}
\end{table}
Fitting this estimate of $\langle\bar{\psi}\psi\rangle$ to the scaling
form over the range $5.41 \leq \beta \leq 5.4225$, used for the lattices of
finite spatial volume gives $\beta_c=5.42270(15)$, $C=1.06(11)$,
$\beta_{mag}=0.24(2)$ at a $78\%$ confidence level. However, it is possible to
fit this extrapolated chiral condensate over the complete range
$5.39 \leq \beta \leq 5.4225$, yielding $\beta_c=5.42267(10)$, $C=1.00(3)$ and
$\beta_{mag}=0.229(9)$ while the confidence level only drops to $60\%$. The plot
of this estimate of the chiral condensate with this latter fit superimposed is
given in figure~\ref{fig:pbp}.
Finally we note that if one fixes $\beta_{mag}=\frac{1}{4}$, its tricritical
value, one obtains a fit with confidence level $89\%$ over the range
$5.41 \leq \beta \leq 5.4225$, which falls to $20\%$ over the complete range
$5.39 \leq \beta \leq 5.4225$. From all these fits we conclude that the value
of $\beta_{mag}$ is consistent with that of the 3-dimensional tricritical
point, but inconsistent with either the $O(4)$ or $O(2)$ values.

\begin{figure}[htb]                                                     
\epsfxsize=6in                                                         
\centerline{\epsffile{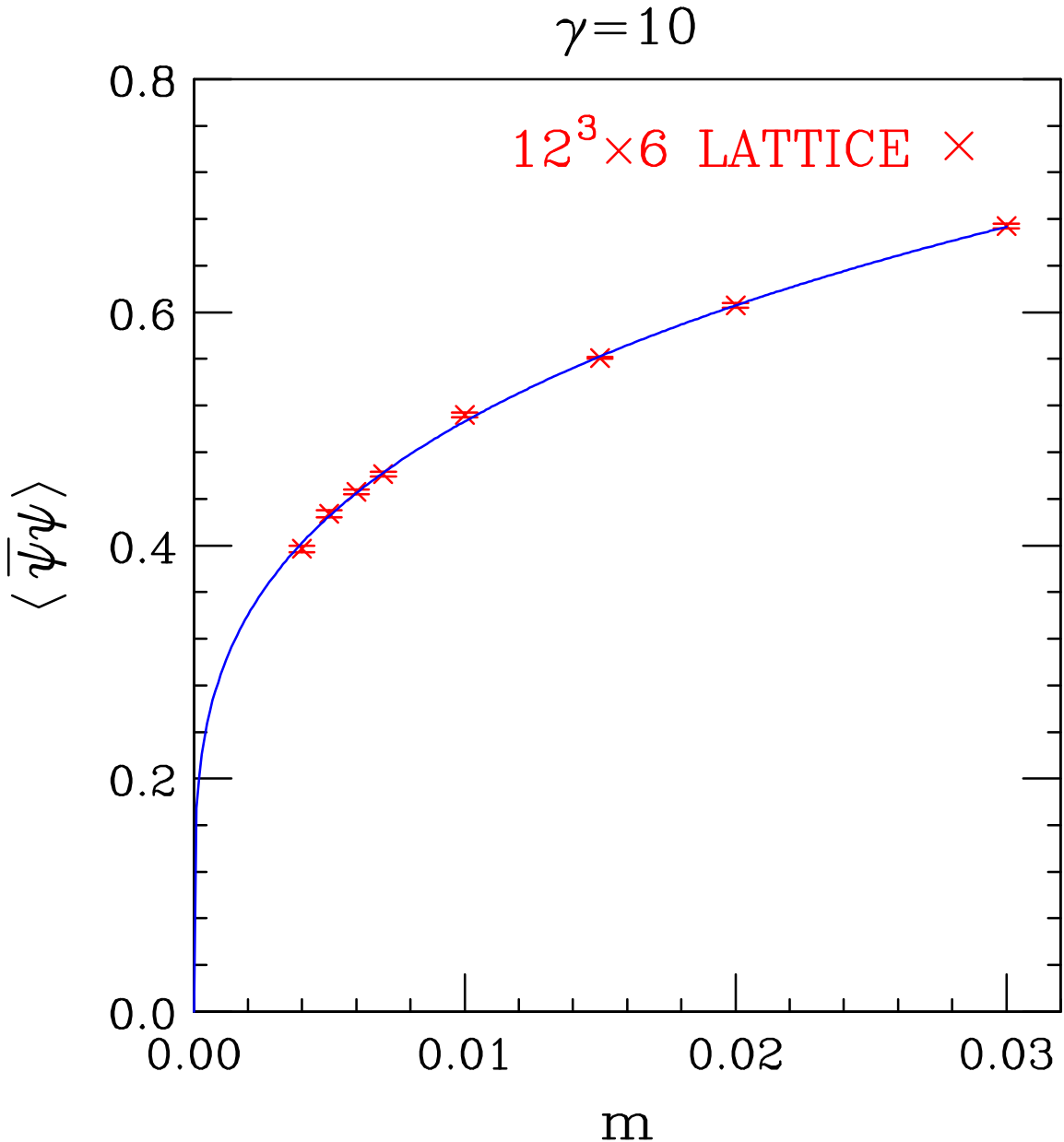}}
\caption{Mass dependence of $\langle\bar{\psi}\psi\rangle$ at $\beta_c$ for
         $\gamma=10$.\label{fig:delta10}}
\end{figure}

We now turn to consideration of the scaling of $\langle\bar{\psi}\psi\rangle$
with quark mass $m$ at $\beta=\beta_c$. This is difficult to perform with
the usual staggered formulation, since it is hard to determine $\beta_c$
with sufficient precision. In addition, for $m$ small enough,
$\langle\bar{\psi}\psi\rangle$ vanishes linearly with $m$ on a configuration
by configuration basis at any $\beta$. In our formulation, not only can
$\beta_c$ be determined with adequate precision, but the chiral condensate
remains finite on each configuration as $m \rightarrow 0$ for $\beta <
\beta_c$ and presumably should vanish as the appropriate power ($1/\delta$) of
$m$ at $\beta=\beta_c$. What a finite mass does is orient the condensate in
the required direction. The value of $m$ required to achieve such orientation
should be somewhat smaller than that required to avoid the region of linear
vanishing in the conventional approach. We fitted the $12^3 \times 6$,
$\gamma=10$ results of table~\ref{tab:pbp-m/10} to the scaling form
$\langle\bar{\psi}\psi\rangle=Dm^{1/\delta}$. Our best fit gives $D=1.66(2)$
and $\delta=3.89(3)$. This fit has only a $4\%$ confidence level. However,
removing the point at $m=0.01$ which lies 2.5 standard deviations off the
curve increases the confidence level to $33\%$ with negligible change to $D$
and $\delta$. We note that this $\delta$ value is incompatible with the $O(4)$
or $O(2)$ values. It does, however, lie between the two $\delta$ values (3 and
5) for the 3-dimensional tricritical point. The more general scaling can be
obtained starting with the $\phi^6$ effective field theory which describes
tricritical behaviour. As argued in section 2, this can be treated in mean
field approximation since it is at the upper critical dimension. The effective
Hamiltonian for this theory is
\begin{equation}
H = \frac{1}{6}\phi^6-mc\frac{1}{5}\phi^5+u\frac{1}{4}\phi^4-mb\frac{1}{3}\phi^3
    +t\frac{1}{2}\phi^2-ma\phi.
\end{equation}
The odd powers of $\phi$ are the symmetry breaking terms. Hence we identify
them with the $m \langle\bar{\psi}\psi\rangle$ term in our Lagrangian, i.e.
\begin{equation}
\langle\bar{\psi}\psi\rangle = a\phi+b\frac{1}{3}\phi^3+c\frac{1}{5}\phi^5.
\label{eqn:pbp-scale}
\end{equation} 
Note here that we have kept the third symmetry breaking operator $\phi^5$ which
has $\delta=1$, since we do not have the luxury of redefining $\phi$ to remove 
it. Since we are considering the case where we are at the tricritical point
for $m=0$, the 2 reduced temperatures $u$ and $t$ are also zero. The vacuum
expectation value of $\phi$ is that which minimizes $H$. Thus $\phi$ is the
real positive solution of
\begin{equation}
\phi^5-m(a+b\phi^2+c\phi^4)=0.
\label{eqn:phi}
\end{equation}
Thus we fit $\langle\bar{\psi}\psi\rangle$ to the form of  
equation~\ref{eqn:pbp-scale} with $\phi$ given in equation~\ref{eqn:phi}. We
obtained a fit with $a=1.05(2)$, $b=2.5(4)$, $c=-3.1(1.4)$. Although this fit
had only a $2\%$ confidence level, removing the point at $m=0.01$ gives a fit
with parameters consistent with those including this point, but with a $35\%$
confidence level. In fact, it is easy to convince oneself that any reasonably
smooth fit would have difficulty fitting this point. We have plotted the
measurements of table~\ref{tab:pbp-m/10} in figure~\ref{fig:delta10} with this
fit superimposed. Estimates of the errors induced in these measurements by
the error in determining $\beta_c$ have been made from the tricritical scaling
function, $f_{QCD}$ and found to be small.

\begin{figure}[htb]                                                           
\epsfxsize=6in                                                   
\centerline{\epsffile{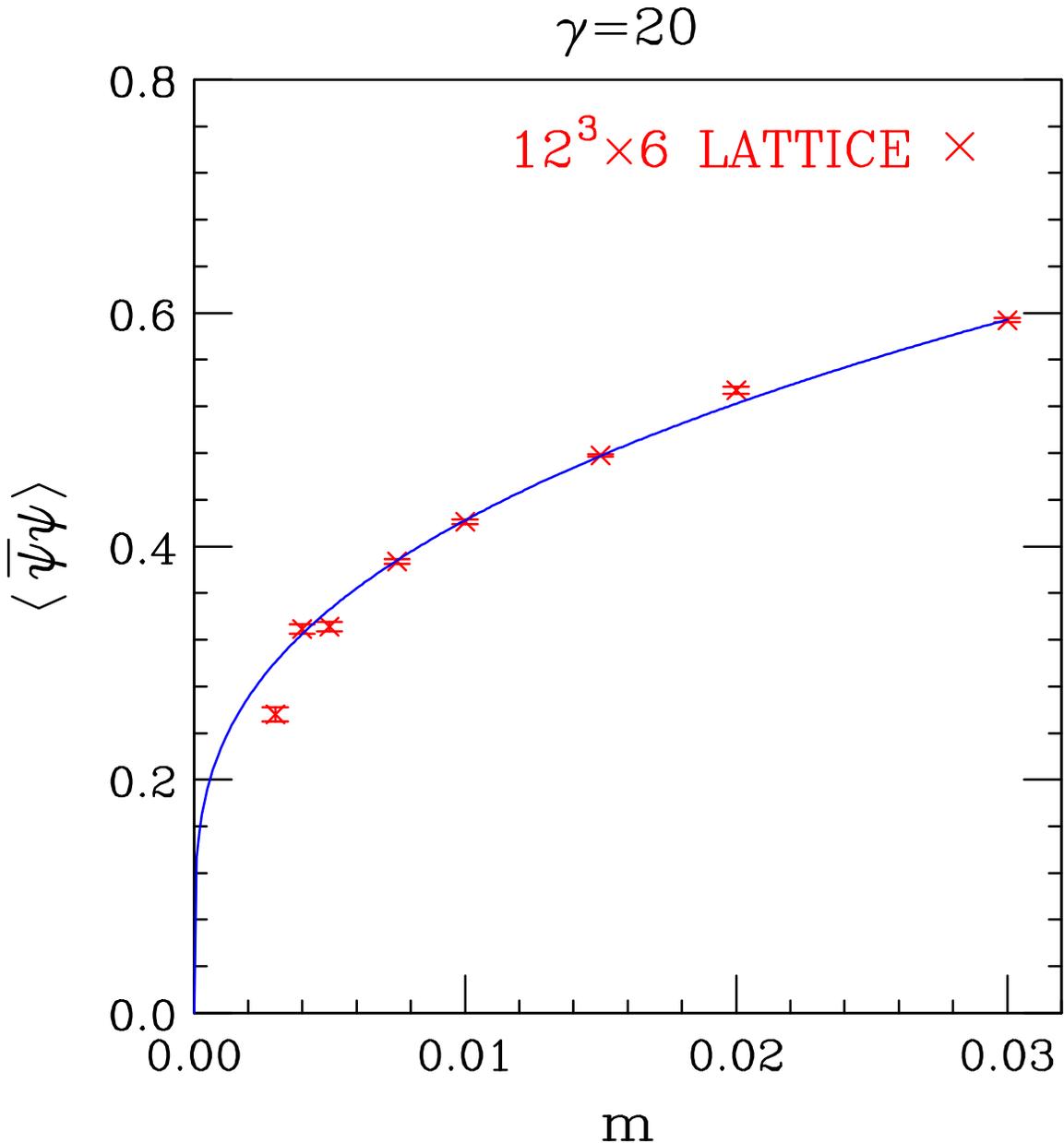}}                           
\caption{Mass dependence of $\langle\bar{\psi}\psi\rangle$ at $\beta_c$ for
         $\gamma=20$.\label{fig:delta20}}                                  
\end{figure}

Our measurements of the mass dependence of $\langle\bar{\psi}\psi\rangle$
on a $12^3 \times 6$ lattice at $\gamma=20$ given in table~\ref{tab:pbp-m/20}
are somewhat poorer than those at $\gamma=10$. For this reason we are unable
to get reasonable fits to this `data'. To clarify this, we have plotted this
`data' in figure~\ref{fig:delta20}.
Just looking at the points, we notice that the $m=0.004$ and $m=0.005$ points
are clearly incompatible. After several trial fits we found it that the 
$m=0.005$ point should be excluded. Smoothness criteria caused us to also
exclude the $m=0.02$ point. Even then no fit came close to the $m=0.003$ point.
Here we suspect that this is because $\beta_c$ was more poorly determined
than in the $\gamma=10$ case. Deviations of $\beta$ from $\beta_c$ have the
most effect on the lowest mass point as is clear from equation~\ref{eqn:eos}.
Excluding these 3 points  allows a fit with $a=0.83(3)$, $b=3.2(5)$,
$c=-2.5(1.5)$ at a $38\%$ confidence level. This curve is plotted in
figure~\ref{fig:delta20}. Clearly such a fit should be considered at best
suggestive, so we base our conclusion that the scaling of
$\langle\bar{\psi}\psi\rangle$ with mass is in accord with tricritical scaling
purely on the $\gamma=10$ measurements.

\begin{figure}[htb]
\epsfxsize=6in                        
\centerline{\epsffile{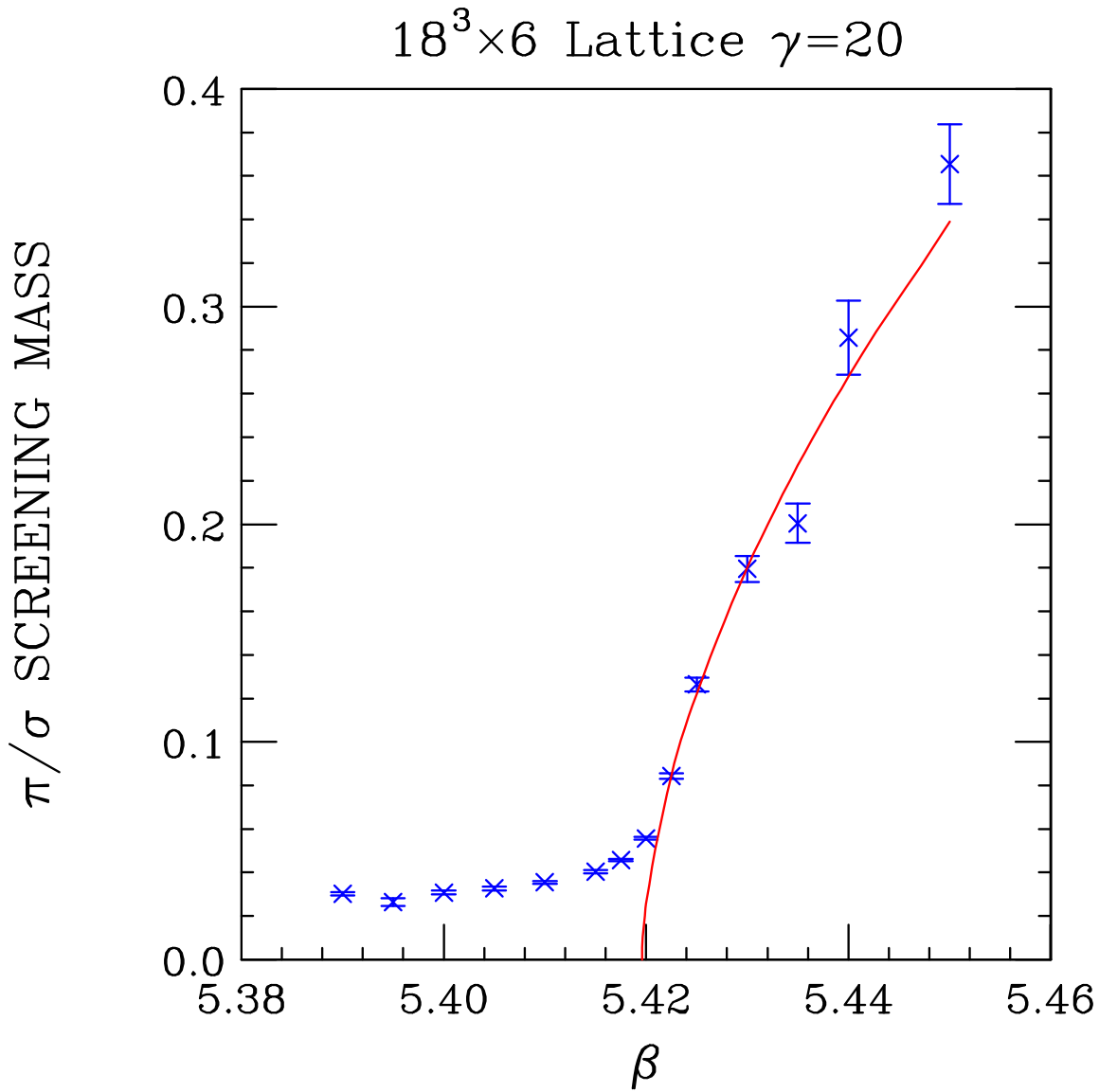}}                                   
\caption{Joint $\sigma$/$\pi$ screening masses as functions of $\beta$.
\label{fig:sigma/pi}}
\end{figure}

From the plot of screening masses of figure~\ref{fig:spmass}, it is clear that
nothing quantitative can be obtained from the $\sigma$ masses in the low
temperature domain. We have therefore concentrated our efforts on fitting the
joint $\pi/\sigma$ screening masses in the plasma region to the scaling form
$m_{\pi/\sigma} = A (\beta-\beta_c)^\nu$. Since different treatment of these
screening masses in the hadronic matter and quark-gluon plasma phases could
bias our results, we chose to adopt the joint fit of the $\sigma$ and $\pi$
propagators of equation~\ref{eqn:ps.mass} over the whole range of $\beta$.
This fit will be correct in the plasma phase. In the low temperature phase,
the non-zero vacuum expectation value of the $\sigma$ field should drive the
fitted `mass' to zero. Fitting these screening masses to the screening form for
$\beta > \beta_c$(in fact we use all points for which $\beta \ge 5.4225$) we
find $A=2.66(71)$, $\beta_c=5.4196(7)$ and $\nu=0.59(7)$ admittedly with only
a $0.3\%$ confidence level. Plotting this fit on a graph
figure~\ref{fig:sigma/pi} of mass measurements, it is clear why: the point at
$\beta=5.35$ is clearly inconsistent with those at $\beta=5.44$ and
$\beta=5.45$. Removing the point at $\beta=5.35$ improves the confidence level
to an acceptable $39\%$ while only changing the fitting parameters within
errors (for example $\nu=0.62(6)$). We attribute this to the systematic errors
in performing fits to point-source propagators with a lattice extent of only
$18$ and limited statistics. From these fits we conclude that $\nu$ is 
consistent with the tricritical value ($\frac{1}{2}$), but does not rule out
an $O(4)$ or $O(2)$ value.

\begin{figure}[htb]
\epsfxsize=6in
\centerline{\epsffile{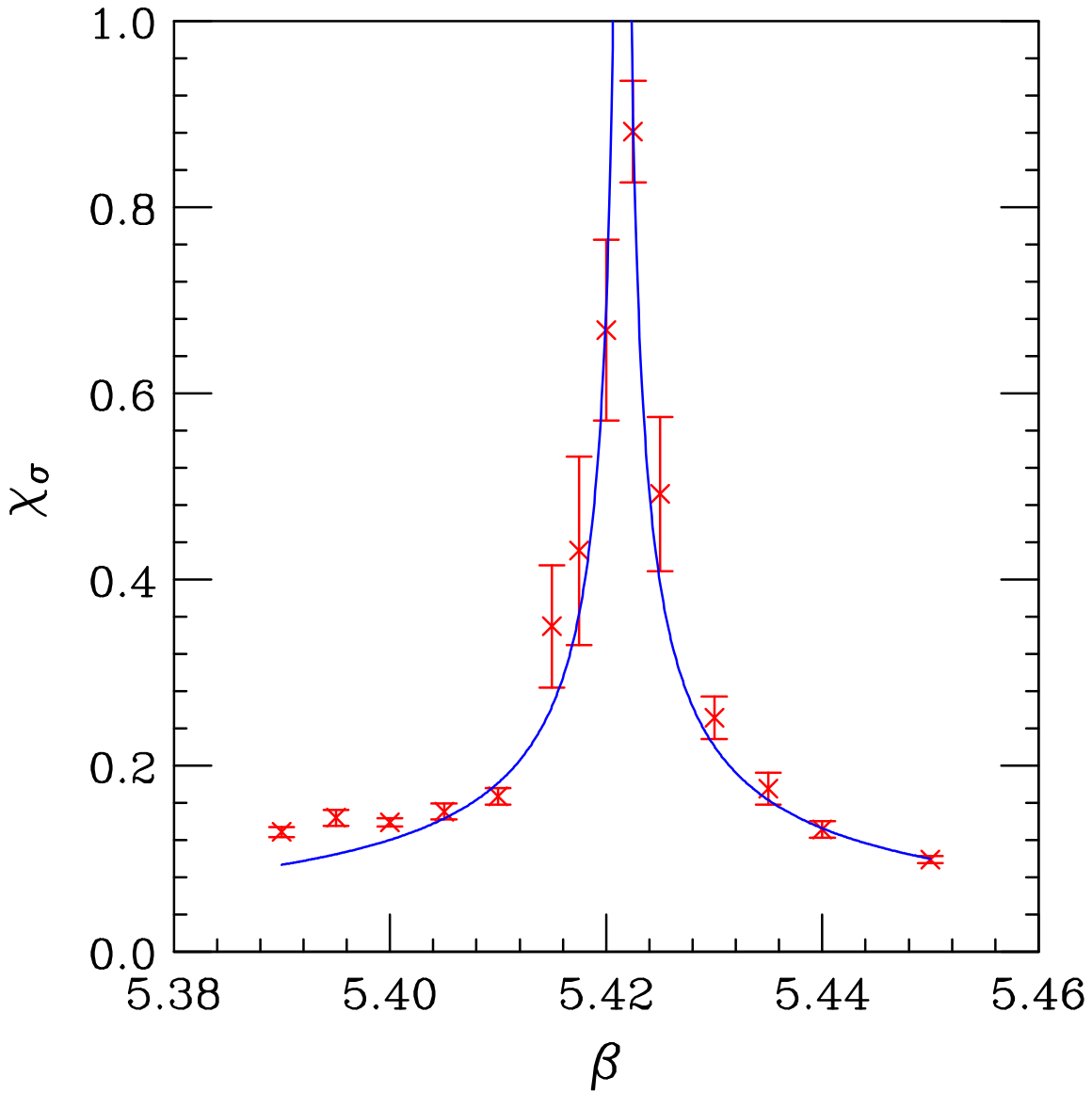}}
\caption{$\sigma$ susceptibility $\chi_\sigma$ as a function of $\beta$.
\label{fig:susc}}
\end{figure}

We have extracted a susceptibility from our $\sigma$ `data' on the 
$18^3 \times 6$ lattice at $\gamma=20$ using
$\chi_\sigma = V \left[\langle\langle\sigma\rangle^2\rangle-                 
\langle\langle\sigma\rangle\rangle^2\right]$. Since our estimate of
$\langle\langle\sigma\rangle\rangle$ was obtained after rotating 
$\langle\sigma\rangle$ to where $\langle\pi\rangle=0$ configuration by
configuration this actually defines a `radial' rather than the conventional
susceptibility. However, this should diverge in the same manner as the
conventional susceptibility, which is all that matters. Fitting to the form
$\chi_\sigma = c |\beta-\beta_c|^{-\gamma_{mag}}$ over the range
$5.405 \le \beta \le 5.45$ gives $\beta_c=5.4215(1)$, $c=0.0096(11)$ and
$\gamma_{mag}=0.66(3)$ with a $31\%$ confidence level. Our measurements and
fit are plotted in figure~\ref{fig:susc}. Since the value of this critical
exponent lies between the 2 $\gamma_{mag}$ values ($\frac{1}{2}$ and $1$)
associated with the tricritical point, but far from the $O(4)$ and $O(2)$
exponents this is further evidence for tricriticality. The reason we did not
use the $\langle\bar{\psi}\psi\rangle$ susceptibility is that our measurements
of $\langle\bar{\psi}\psi\rangle$ are noisy estimators using a single noise
vector per configuration which gives extra contributions to the susceptibility
estimate that can only be removed by use of more than one noise vector per 
configuration.

\section{Discussion and Conclusions}

We have presented evidence that the $N_t=6$ finite temperature phase
transition for 2-flavour staggered lattice QCD with extra chiral 4-fermion
interactions is in the universality class of a 3-dimensional tricritical point
rather than that of the 3-dimensional $O(4)$ or $O(2)$ sigma model, which was
expected. Since a tricritical point occurs when there are extra relevant
operators with consequent competing transitions, we believe this to be due to
the extra higher-dimensional operators which can be important on such coarse
lattices. For this reason, we believe the $O(4)$/$O(2)$ universality class
would be observed for sufficiently fine lattices, i.e. those with sufficiently
large $N_t$. Since the $N_t=4$ transition with $\gamma=10$ was observed to be
first order, and a tricritical point marks the transition from first order to
universal second order behaviour, this strongly suggests that $O(4)$/$O(2)$
universality would be observed for $N_t$ as small as $8$.

The fact that our modified action allowed us to work at zero quark mass was
crucial in allowing us to come to these conclusions. This enabled us to measure
the critical exponent $\beta_{mag}$ directly. Our best estimates lay in the
range $0.23 \leq \beta_{mag} \leq 0.27$ (with errors typically around 2 in the
second decimal place) which is consistent with the $\frac{1}{4}$ of the
tricritical point, but inconsistent with the $0.384(5)$ of the $O(4)$ sigma
model, the $0.35(1)$ of the $O(2)$ sigma model and the $\frac{1}{2}$ of mean
field theory. Because these scaling fits give precise estimates for $\beta_c$,
it is possible to measure $\delta$ directly from the scaling of the chiral
condensate with quark mass with $\beta$ fixed at $\beta_c$. This value,
$\delta=3.89(3)$, lay between the two $\delta$ values (3 and 5) of the
tricritical point, and a satisfactory fit to the more complex scaling this
implies was obtained. Again this behaviour was incompatible with $O(4)$/$O(2)$
scaling where $\delta \approx 4.8$ and mean field scaling with $\delta=3$. The
scaling of the $\sigma/\pi$ mass in the plasma domain yielded a $\nu$ value
which was compatible with tricritical, $O(4)$/$O(2)$ and mean field
scaling. Finally we were able to extract the susceptibility critical exponent
$\gamma_{mag}=0.66(3)$ which lies between the two values ($\frac{1}{2}$ and
$1$) of the tricritical point but disagrees with the $O(4)$/$O(2)$ values
($1.328(6)$ and $1.471(6)$) and the mean field value ($1$).

Since we believe we are seeing more complex phase structure which is driven by
lattice artifacts, the behaviour we are seeing does not have to apply to the
standard staggered action. However, since scaling analyses with the standard
action on $N_t=4$ lattices showed definite departures from $O(4)$/$O(2)$
universality \cite{aoki,boyd,milc,laermann}
it strongly suggests that some such behaviour is present. In fact,
based on our earlier, $N_t=4$ results one of the collaborations suggested 
\cite{milc} that
these anomalies might be associated with the first order transition we reported
in \cite{lks}.
Unfortunately there are not yet sufficient high precision measurements with
the standard action at $N_t=6$ to tell if it too displays departures from
$O(4)$/$O(2)$ universality.

Clearly, simulations at $N_t=8$ should be next on our agenda. We would also
suggest that the addition of such 4-fermion terms should be considered for
other actions, especially improved staggered actions.

\section*{Acknowledgements}

These computations were performed on the CRAY C90/J90/SV1's and T3E at NERSC,
and on a T3E at SGI/Cray. We thank J.-F.~Laga\"{e} who contributed to the
earlier stages of this work. We would also like to thank M.~Stephanov for
educating us on tricritical behaviour. Finally we wish to thank C.~DeTar for
providing us with the $O(4)$ scaling functions used by the MILC collaboration.
This work was supported by the U.~S. Department of Energy under contract
W-31-109-ENG-38 and the National Science Foundation grant NSF-PHY96-05199.

\end{document}